\documentclass[letterpaper,12pt]{article}
\pdfoutput=1
\usepackage{lineno}  

\usepackage{amsfonts}
\usepackage{amssymb}
\usepackage{amsmath}
\usepackage{xspace}
\usepackage{bm}
\usepackage{xfrac}

\usepackage{alltt}
\usepackage{latexsym}
\usepackage{lineno}
\usepackage[]{natbib}
\bibpunct{(}{)}{,}{a}{}{,}

\usepackage{url}
\usepackage{xspace}
\usepackage{algpseudocode}
\usepackage{color}
\usepackage[nofiglist]{endfloat}
\usepackage{ulem}

\usepackage{setspace}
\newcommand{\Link}{L2010\xspace}

\newtheorem{mythm}{Theorem}

\newcommand{\Mt}{M$_\text{t}$\xspace}
\newcommand{\Mh}{M$_\text{h}$\xspace}
\newcommand{\Mb}{M$_\text{b}$\xspace}
\newcommand{\Mth}{M$_\text{th}$\xspace}

\newcommand{\Mta}{M$_{\text{t}\alpha}$\xspace}

\newcommand{\mrsadd}[1]{\textcolor{black}{#1}}

\begin{document}

\allowdisplaybreaks

\title{Connecting the Latent Multinomial}
\author{Matthew R. Schofield$^{1}$\thanks{E-mail: \texttt{mschofield@maths.otago.ac.nz}}, Simon J. Bonner$^{2}$\\
  \normalsize{$^{1}$Department of Mathematics and Statistics, University of Otago, New Zealand.}\\
  \normalsize{$^{2}$Department of Statistics, University of Kentucky, Lexington, KY, USA.}\\
} 
\date{}
\maketitle

\begin{abstract}
  \cite{Link2010a} define a general framework for analyzing capture-recapture data with potential misidentifications. In this framework, the observed vector of counts, $\bm y$, is considered as a linear function of a vector of latent counts, $\bm x$, such that $\bm y = \bm A \bm x$, with $\bm x$ assumed to follow a multinomial distribution conditional on the model parameters, $\bm \theta$. Bayesian methods are then applied by sampling from the joint posterior distribution of both $\bm x$ and $\bm \theta$. In particular, \cite{Link2010a} propose a Metropolis-Hastings algorithm to sample from the full conditional distribution of $\bm x$, where new proposals are generated by sequentially adding elements from a basis of the null space (kernel) of $\bm A$. 
We consider this algorithm and show that using elements from a simple basis for the kernel of $\bm A$ may not produce an irreducible Markov chain. Instead, 
we require a Markov basis, as defined by \cite{Diaconis1998}.  We illustrate the importance of Markov bases with three capture-recapture examples.  
We prove that a \mrsadd{specific} lattice basis is \mrsadd{a Markov basis} for a class of models including the original model considered by \cite{Link2010a} and confirm that the specific basis used by \cite{Link2010a} for their example with two sampling occasions is a Markov basis.  The constructive nature of our proof provides an immediate method to obtain \mrsadd{a Markov basis} for \mrsadd{any model} in this class.   
\end{abstract}

\section{Introduction}\label{sect:intro}
The development of capture-recapture methodology has a long history, allowing estimation of demographic parameters of interest for \mrsadd{animal} populations \cite[see][for a review]{Amstrup2005}. Similar methods have also been used to study human populations, including intravenous drug users \cite[]{King2009} and human rights \mrsadd{abuse} victims \cite[]{Lum2013}. In general, a capture-recapture experiment consists of a series of capture occasions on which overlapping subsets of the population are observed. For animal populations the occasions are usually ordered in time while for human populations they may comprise lists obtained from different sources. It is assumed that each individual has a unique identifying mark that is either given or realized when the individual is first captured and this mark can be used to identify the individual on subsequent occasions.
In this paper, we are concerned with fitting capture-recapture models to data that provide an incomplete or inaccurate representation of the true encounters of individuals during the experiment. This may occur if the data consist of incomplete summary statistics or if individuals are misidentified on some occasions.
Examples of capture-recapture studies that are prone to identification errors include (i) multi-list studies in which individuals may be matched based on personal information such as name, birth date, medical record number \cite[]{Seber2000,Lee2001,Sutherland2005,Fienberg2009}, (ii) animal studies in which individual identity is found from non-invasive sampling, e.g. genetic information from scat or hair \cite[]{Wright2009,Link2010a,Yoshizaki2011} or photographic ID of individuals \cite[]{Yoshizaki2009, Bonner2013a, McClintock2013}, and (iii) studies in which (at least) two sources of capture-recapture information are available for the same population with little to no information about how the individual IDs in one source corresponds to individual ID from the other sources \cite[]{Bonner2013a,McClintock2013}.

Our focus is on the algorithm for a general class of mark-recapture models allowing for misidentification considered by \cite{Link2010a} (hereafter \Link).  This class is described by the latent multinomial model, in which an observed data vector, $\bm y$ can be expressed as a linear function of a latent data vector, $\bm x$, modeled by a multinomial distribution \mrsadd{with unknown parameters $\bm \theta$, denoted $[\bm x|\bm \theta]$.  The notation $[x]$ denotes the probability density function $f_{X}(x)$ for a continuous random variable $X$ or the probability mass function $\Pr(X=x)$ for a discrete random variable $X$.} The linear function is expressed as
\begin{linenomath*}
\begin{equation}\label{eq:lincon}
\bm y = \bm A \bm x,
\end{equation}
\end{linenomath*}
where $\bm A$ is called the configuration matrix (a matrix of known constants that depends on the specific problem) \mrsadd{with more columns than rows}.  
We continue to call this modeling setup the latent multinomial model, \mrsadd{even though the setup is flexible and can accommodate other probability mass functions $[\bm x|\bm \theta]$, such as the Poisson model considered by \cite{Lee2002}}.

The goal is to sample from the joint posterior distribution $[\bm \theta,\bm x|\bm y]$ using Markov chain Monte Carlo (MCMC) by alternating between sampling from the full conditional distributions $[\bm \theta|\bm x, \bm y]$ and $[\bm x|\bm y, \bm \theta]$.  The difficulty with this approach is in specifying an updating scheme for $\bm x$.  That is, how to \mrsadd{efficiently} sample from $[\bm x|\bm y,\bm \theta]$ in such a way so that every $\bm x$ vector that satisfies (\ref{eq:lincon}) has a positive probability of being reached at some point during the updating.  We consider three examples demonstrating that the scheme for updating $\bm x$ proposed by \Link may not produce an irreducible Markov chain for models within the latent multinomial framework.  We then present theory identifying a class of models for which the specific algorithm  does produce irreducible Markov chains, and show more generally how these methods fit within the framework of algebraic statistics. This allows us to develop an extension of the algorithm which can be used to generate valid MCMC samplers for the posterior distributions from a broader class of latent multinomial models.

The MCMC algorithm we consider throughout this manuscript is presented in Figure \ref{fig:alg}.
Starting with an initial state $\bm x^{0}$ satisfying the linear constraint, a proposal is generated on the first iteration by adding or subtracting an element chosen randomly from a
subset of the kernel (or null space) of $\bm A$, $\mathcal B=\{\bm a_{1}, \bm a_{2}, \ldots, \bm a_{m}\} \subset \ker(\bm A) $, with cardinality $m$. The proposal is then accepted or rejected with probability determined by the Hasting's ratio, $r$, and the algorithm continues to the second iteration.
This algorithm is a modification of that presented by \Link, with three differences: (i) \Link steps through all $m$ elements in $\mathcal B$ in order instead of selecting an element at random on each iteration, (ii) when stepping through every element in $\mathcal B$, \Link multiplies element $\bm a_{i}$ by a coefficient $c \in \{-C_{i},\ldots,-1,1,\ldots,C_{i}\}$ in order to improve convergence, and (iii) 
\mrsadd{\Link assumes that $\mathcal B$ is a basis for $\ker(\bm A)$, while we allow $\mathcal B$ to be a more general subset that spans $\ker(\bm A)$.}
The first two differences may impact the efficiency of the algorithm but do not change the stationary distribution of the resulting Markov chains, and we do not consider these differences further. Our focus is on the third difference and the effect that the set $\mathcal B$ can have on the generated Markov chains and their stationary distributions. 

\begin{figure}
\centering
\begin{algorithmic}[1]
\State Initialize $\bm x^{0}$ so that $\bm y = \bm A \bm x^{0}$
\For{$i = 1:n$} 
\State Sample $k \in \{1,2,\ldots,m\}$ with equal probability
\State Sample $c \in \{-1,1\}$ with equal probability
\State Set $\bm x_{\mathrm{cand}} = \bm x^{i-1} + c\bm a_{k}$
\State Calculate the metropolis acceptance probability: $r = \min\left(1,\frac{[\bm x_{\mathrm{cand}}|\theta]}{[\bm x^{i-1}|\theta]}\right)$
\State Accept $\bm x_{\mathrm{cand}}$ with probability $r$ (if accepted $\bm x^{i}=\bm x_{\mathrm{cand}}$; otherwise $\bm x^{i}=\bm x^{i-1}$)
\EndFor
\end{algorithmic}
\caption{Algorithm for updating the latent counts $\bm x$.  The value $n$ is the number of iterations in the algorithm and the vectors $\mathcal B = \{\bm a_{1}, \bm a_{2}, \ldots, \bm a_{m}\}$ are a subset of the kernel of $\bm A$.}
\label{fig:alg}
\end{figure}

\begin{figure}[!htbp]
\centering
\includegraphics[width=0.7\textwidth]{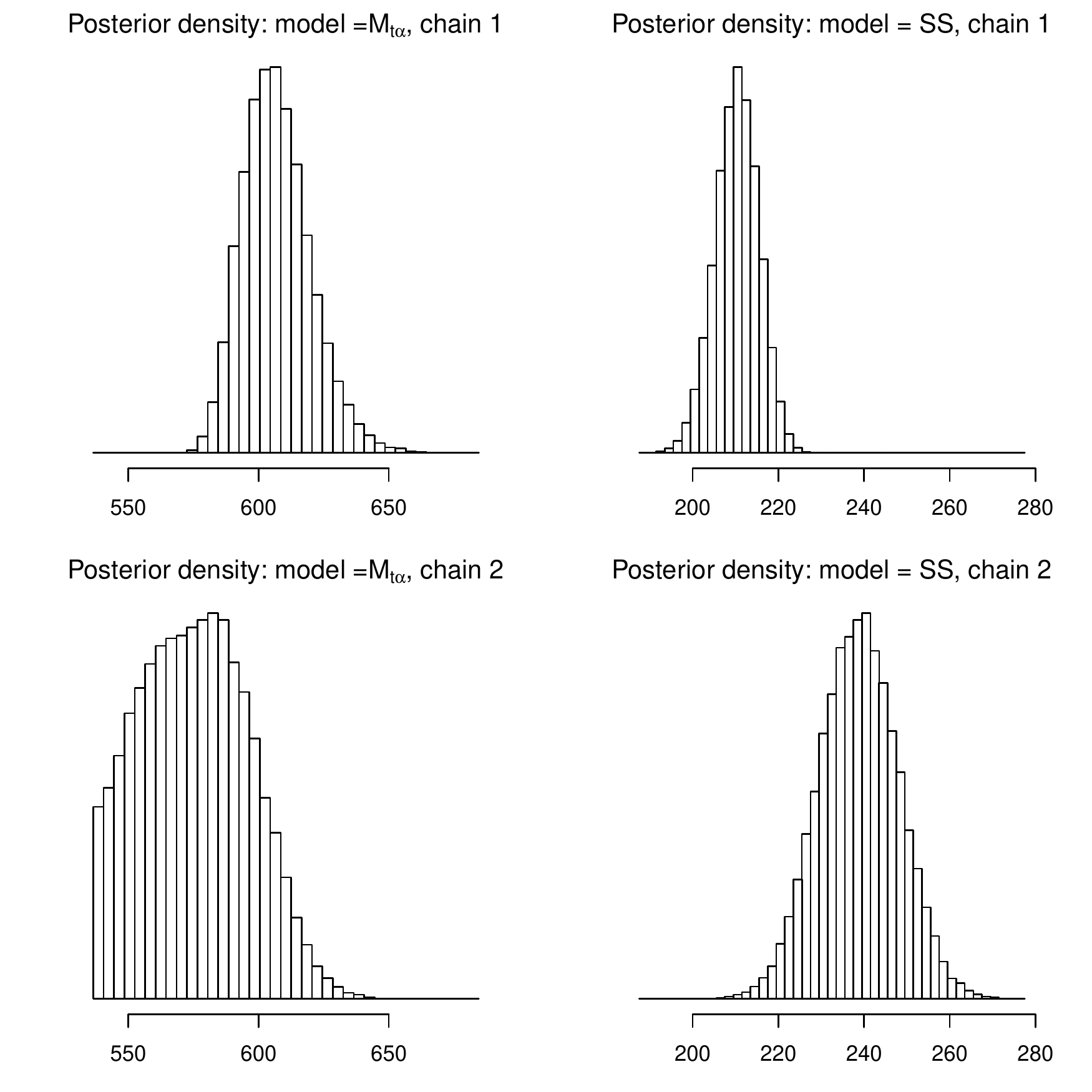}
\caption{Estimated posterior densities of a quantity of interest for model \Mta (left panel) and a multi-list model where summary statistics are presented in place of full data (SS; right panel).  Within each model, the resulting density estimates are plotted separately from the output of two parallel MCMC algorithms (for each model) with different starting values.}
\label{fig:twostart}
\end{figure}

To illustrate the problems that may occur if $\mathcal B$ is poorly specified we consider three examples of models which fit into the latent multinomial framework. First we consider the same closed population mark-recapture model with misidentification considered by \Link. This model, called \Mta, assumes that captures occur according to a closed population model with time dependent capture probabilities and that errors in identifying an individual are unique and create ghost histories with single captures. Second, we consider a multi-list modeling problem in which summary statistics are presented in place of the full data set, possibly for privacy reasons.  Our aim is to sample from possible complete data sets with the given sufficient statistics. Finally, we consider a more complicated model of misidentification in mark-recapture which allows for one marked individual to be identified as another previously marked individual. Full details of these models and the issues regarding the selection of the \mrsadd{set} $\mathcal B$ to be used in the algorithm in Figure \ref{fig:alg} are provided in sections \ref{sect:mta}, \ref{sect:summary}, and \ref{sect:resight}. As motivation, we consider the output from Markov chains constructed using the algorithm in Figure \ref{fig:alg} for each of the three examples. For each example, \mrsadd{we defined $\mathcal B$ to be a basis for $\ker(\bm A)$ as in \Link and ran two parallel chains, each of which started from a different initial value.}  For both model \Mta and the multi-list model with sufficient statistics, despite strong evidence that each chain has converged, it is clear that the two chains are not sampling from the same distribution for a given quantity of interest (Figure \ref{fig:twostart}).  This is even more apparent in the third example where one of the two chains never moves from its initial value.

The problem in all three examples is that the stationary distribution reached by the Markov chains produced by the algorithm in Figure \ref{fig:alg} may depend on the chosen \mrsadd{set}, $\mathcal B$ \mrsadd{and the initial value of $\bm x$}. Although the values of $\bm x$ proposed on each iteration are guaranteed to satisfy the linear constraint the resulting Markov chains may not reach all points in the sample space and the stationary distributions may be dependent on the initial values. In the next section we provide a basic introduction to the field of algebraic statistics and the results of \cite{Diaconis1998} and others who have explored approaches for sampling from $\bm x$ from a linear constraint as in (\ref{eq:lincon}) in other application areas.  We then consider the implications of this theory \mrsadd{to show why the MCMC algorithms failed above (Figure \ref{fig:twostart}), and how} valid MCMC samplers can be constructed for each of the three examples.
\section{Introduction to algebraic statistics}\label{sect:basis}

Sampling $\bm x$ in the presence \mrsadd{of} the linear constraint in (\ref{eq:lincon}) is not unique to capture-recapture problems.  In a seminal paper in algebraic statistics, \cite{Diaconis1998} considered a linear constraint of the same form when developing conditional goodness-of-fit  tests  for contingency tables \cite[see][for a recent review]{Karwa2013}.  That is, they considered how to construct an MCMC algorithm to sample different contingency tables with common (fixed) row and column sums (such ideas can also be extended to multi-way contingency tables).

To consider the problem at hand in more detail we will summarize several definitions and results from linear algebra in this section (basic definitions regarding kernels and bases are provided in the supplementary materials).  We will use a $3 \times 3$ contingency table example to illustrate many of the ideas.  The table is
\begin{linenomath*}
\[
\begin{array}{rrr|r}
x_{11} & x_{12} & x_{13} & x_{1\cdot}\\
x_{21} & x_{22} & x_{23} & x_{2\cdot}\\
x_{31} & x_{32} & x_{33} & x_{3\cdot}\\\hline
x_{\cdot1} & x_{\cdot2} & x_{\cdot3} &
\end{array}
\]
\end{linenomath*}
where $x_{ij}$ is the value in the $i$th row and $j$th column, $x_{\cdot j}$ refers to the sum of the $j$th column and $x_{i \cdot}$ refers to the sum of the $i$th row.
The column and row sums are vectorized to give the vector of \mrsadd{summary} statistics
\begin{linenomath*}
\[
\bm y = (x_{\cdot1},x_{\cdot2},x_{\cdot3},x_{1\cdot},x_{2\cdot})^{\prime}.
\]
\end{linenomath*}
Note that we need not include the third row sum as this is a derived quantity of the other elements of $\bm y$.  The individual entries in the table are vectorized to give
\begin{linenomath*}
\[
\bm x = (x_{11},x_{21},x_{31},x_{12},x_{22},x_{32}, x_{13}, x_{23}, x_{33})^{\prime}.
\]
\end{linenomath*}
The specification is completed with
\begin{linenomath*}
\[
 \bm A = \left(\begin{array}{ccccccccc}
 1 & 1 & 1 & 0 & 0 & 0 & 0 & 0 & 0\\
 0 & 0 & 0 & 1 & 1 & 1 & 0 & 0 & 0\\
 0 & 0 & 0 & 0 & 0 & 0 & 1 & 1 & 1 \\
 1 & 0 & 0 & 1 & 0 & 0 & 1 & 0 & 0 \\
 0 & 1 & 0 & 0 & 1 & 0 & 0 & 1 & 0
 \end{array}\right)
 \]
\end{linenomath*}
so that the constraints inherent in a contingency table follow (\ref{eq:lincon}).
If we have column/row sums given by
\begin{linenomath*}
\[
\bm y = (5, 3, 2, 0, 4)^{\prime}
\]
\end{linenomath*}
then two contingency tables compatible with these constraints have entries
\begin{linenomath*}
\begin{equation}\label{eq:twosol}
\bm x_1 = (0,2,3,0,1,2,0,1,1)^{\prime}~~~\text{and}~~~
\bm x_{2} = (0,3,2,0,0,3,0,1,1)^{\prime}.
\end{equation}
\end{linenomath*}

Our goal is to specify an MCMC algorithm that samples from the set of vectors $\bm x$ that satisfy (\ref{eq:lincon}) for a particular $\bm y$. 
This is defined as
the $\bm y$-fiber (or simply fiber) $\mathcal F_{\bm y}$,
\begin{linenomath*}
\[
\mathcal F_{\bm y}=\{\bm x \in \mathbb N^{d}: \bm y = \bm A \bm x\},
\]
\end{linenomath*}
where $d$ is the dimension of $\bm x$ and $\mathbb N = \{0,1,\ldots\}$.  \mrsadd{\Link refers to $\mathcal F_{\bm y}$ as the feasible set.}

To move between elements of the fiber, we make use of the lattice kernel $\ker_{\mathbb Z}(\bm A)$.  
The lattice kernel is the integer valued subset of the kernel,
\begin{linenomath*}
\[\ker_{\mathbb Z}(\bm A)=\ker(\bm A) \bigcap \mathbb Z^{d}=\{\bm x \in \mathbb Z^d: \bm A \bm x=\bm 0\}.\]
\end{linenomath*}
\mrsadd{In algebraic statistics, a move is defined to} be any element of the lattice kernel, such that the vector $\bm v$ is a move if $\bm v \in \ker_{\mathbb Z}(\bm A)$.
An implication of this is that 
if $\bm x_1,\bm x_2 \in \mathcal F_{\bm y}$ then $\bm x_2 - \bm x_1$ is a move.
\mrsadd{The idea is that the elements of the lattice kernel can be added to a vector that satisfies the linear constraint and the result is guaranteed to still satisfy the constraint.  However, it is not practical to consider all elements of the lattice kernel when updating $\bm x$ as $\ker(\bm A)$ is potentially very large and difficult to compute.  Instead we want to find a smaller
}
set of moves $\mrsadd{\mathcal B} = \{\bm v_1,\ldots,\bm v_m\} \subset \ker_{\mathbb Z}(\bm A)$ that can be used 
to update $\bm x$.  
\mrsadd{That is, we require a smaller set of moves so that it is possible to move between all elements of $\mathcal F_{\bm y}$ using the algorithm in Figure \ref{fig:alg}.}

The suggestion of \Link was to use a basis for $\ker(\bm A)$ \mrsadd{for} this set of moves.  However, we do not wish to construct a basis for $\ker(\bm A)$, but \mrsadd{instead a lattice basis for} the integer lattice $\ker_{\mathbb Z}(\bm A)$\mrsadd{.  A lattice basis is a set of linearly independent vectors where every $\bm v \in \ker_{\mathbb Z}(\bm A)$ can be found as a linear combination of the lattice basis vectors using integer coefficients.}   If we insist on \mrsadd{using a basis for $\ker(\bm A)$}, it may not be possible to reach all solutions using only integer values of the coefficients, $c$, as specified in the algorithm in Figure \ref{fig:alg}.  
\mrsadd{However,} even if we choose \mrsadd{to use} a lattice basis for \mrsadd{$\mathcal B$ it may be necessary to pass through one (or more) vectors containing negative elements when applying moves one at a time to transition between elements in the fiber $\mathcal F_{\bm y}$.  As vectors $\bm x$ containing negative elements can never be accepted, the use of a lattice basis for $\mathcal B$ may result in sampling from a subset of the fiber $\mathcal F_{\bm y}$ when using} the algorithm in Figure \ref{fig:alg}.  This explains the observed results in the three examples shown in Section \ref{sect:intro}: the two chains are exploring different subsets of the fiber.


\mrsadd{These ideas are} formalized using the concept of connectivity.
Elements $\bm x_j, \bm x_k \in \mathcal F_{\bm y}$ are connected using the set $\bm V = (\bm v_1,\ldots,\bm v_m)$ if there are moves $\bm v_{i} \in \bm V, ~i \in \{1,\ldots,M\}$ so that we can start from $\bm x_j$ and add or subtract these moves one at a time to reach $\bm x_k$ without any element in any of the partial sums ever being negative (note that the elements $\bm v_i, ~i=1,\ldots,M$ need not be distinct and some elements may be repeated multiple times).  That is, there exist $\epsilon_{1},\ldots,\epsilon_{M} \in \{-1,1\}$ such that
\begin{linenomath*}
\[
\bm x_k = \bm x_j + \sum_{j=1}^{M}\epsilon_{j}\bm v_{j} \hspace{1cm} \text{and} \hspace{1cm} \bm x_{1} + \sum_{k=1}^{L} \epsilon_{k}\bm v_{k} \in \mathcal F_{\bm y}, ~~L = 1,\ldots,M-1.
\]
\end{linenomath*}
We then say that the fiber $\mathcal F_{\bm y}$ is connected \mrsadd{by $\bm V$} if every pair of elements in the fiber are connected.

We can apply the algorithm in Figure \ref{fig:alg} to the $3 \times 3$ contingency table example using the elements of \mrsadd{a} lattice basis.  A lattice basis can be found using the Hermite normal form \cite[][pg. 53]{Aoki2012}.
Unless otherwise stated, all lattice bases provided in this manuscript are found using this approach.  We note that the lattice basis obtained is not unique and a different basis is often found if one reorders the columns of $\bm A$ (and corresponding entries of $\bm x$).
For the contingency table, \mrsadd{a lattice} basis is given by elements LB1 -- LB4 in (\ref{eq:KBcont})
\begin{linenomath*}
\begin{equation}\label{eq:KBcont}
\arraycolsep=0.7em
\def\arraystretch{0.7}
\begin{array}{r|rrrrrrrrr}
& x_{11} & x_{21} & x_{31} & x_{12} & x_{22} & x_{32} & x_{13} & x_{23} & x_{33}\\\hline
\text{LB1}	&	1	&	-1	&	0	&	-1	&	1	&	0	&	0	&	0	&	0	\\
\text{LB2}	&	-1	&	0	&	1	&	1	&	0	&	-1	&	0	&	0	&	0	\\
\text{LB3}	&	1	&	-1	&	0	&	0	&	0	&	0	&	-1	&	1	&	0	\\
\text{LB4}	&	0	&	0	&	0	&	1	&	0	&	-1	&	-1	&	0	&	1
\end{array}
\end{equation}
\end{linenomath*}
If we attempt to apply any of the elements LB1 --- LB4 to either $\bm x_1$ or $\bm x_2$ in (\ref{eq:twosol}) we immediately find a problem.  Either adding or subtracting any of LB1 -- LB4 results in at least one negative count in the proposal and will lead to it being automatically rejected.  That means there is no way to use the elements LB1 -- LB4 as moves in the algorithm in Figure \ref{fig:alg} and successfully transition between the two solutions in (\ref{eq:twosol}).  In fact, we are unable to move between any two valid solutions.  As a result, the lattice basis in (\ref{eq:KBcont}) does not connect the fiber for this example.  One solution is to change the algorithm in Figure \ref{fig:alg} to use elements of a lattice basis in a linear combination instead of one-at-a-time.  While attractively simple, \cite{Diaconis1998} implemented this for several examples and found that it was inefficient and did not work well in practice. We do not consider this further.

To overcome the shortcomings of constructing moves via integer multiples of an element from a lattice basis, we take a Markov basis for the set $\mathcal B$ \cite[]{Diaconis1998}.  A Markov basis is a larger set of elements in $\ker_{\mathbb Z}(\bm A)$ that connects all fibers $\mathcal F_{\bm y}$ irrespective of the given values in $\bm y$.
%
A finite set $\mathcal M \subset \ker_{\mathbb Z}(\bm A)$ is a Markov basis if, for any $\bm y$ such that $\mathcal F_{\bm y} \neq \varnothing$ and for all elements $\bm x_1, \bm x_2 \in \mathcal F_{\bm y}$, $\bm x_1 \neq \bm x_2$, there exist $M > 0$, $\bm v_{1},\ldots,\bm v_{M} \in \mathcal M$ and $\epsilon_{1},\ldots,\epsilon_{M} \in \{-1,1\}$ such that
\begin{linenomath*}
\[
\bm x_2 = \bm x_1 + \sum_{j=1}^{M}\epsilon_{j}\bm v_{j} \hspace{1cm} \text{and} \hspace{1cm} \bm x_{1} + \sum_{k=1}^{L} \epsilon_{k}\bm v_{k} \in \mathcal F_{\bm y}, ~~L = 1,\ldots,M-1.
\]
\end{linenomath*}
The first condition says that we can use moves from a Markov basis as in the algorithm in Figure \ref{fig:alg} to move between any two elements of our fiber.  The second condition says that when moving between any two elements in the fiber, we always remain in the fiber (i.e. we never encounter a negative count).  

Although Markov bases are relatively easy to describe there is no simple algorithm for their computation. \citet{Diaconis1998} show how a Markov basis can be computed using techniques from commutative
algebra. The theory is based on
what is now known as the Fundamental Theorem of Markov Bases which describes how finding a Markov basis is equivalent to finding a set of generators of a toric ideal in a polynomial ring associated with the matrix $\bm A$.  We refer the interested reader to
\citet{Cox2007} for details on commutative algebra and to
\citet{Diaconis1998}, \cite{Drton2009}, \citet{Aoki2012} and the references therein for additional information on the generation of Markov bases in algebraic statistics.
Unless otherwise stated, we use the freely available software {\tt 4ti2} \cite[]{Hemmecke2013} to compute the Markov bases for the examples in this manuscript.

For the $3\times3$ contingency table, a Markov basis consists of the nine elements in (\ref{eq:MBcont})
\begin{linenomath*}
\begin{equation}\label{eq:MBcont}
\arraycolsep=0.7em
\def\arraystretch{0.7}
\begin{array}{r|rrrrrrrrr}
& x_{11} & x_{21} & x_{31} & x_{12} & x_{22} & x_{32} & x_{13} & x_{23} & x_{33}\\\hline
\text{MB1}	&	0	&	0	&	0	&	0	&	1	&	-1	&	0	&	-1	&	1	\\
\text{MB2}	&	0	&	0	&	0	&	1	&	-1	&	0	&	-1	&	1	&	0	\\
\text{MB3}	&	0	&	0	&	0	&	1	&	0	&	-1	&	-1	&	0	&	1	\\
\text{MB4}	&	0	&	1	&	-1	&	0	&	-1	&	1	&	0	&	0	&	0	\\
\text{MB5}	&	0	&	1	&	-1	&	0	&	0	&	0	&	0	&	-1	&	1	\\
\text{MB6}	&	1	&	-1	&	0	&	-1	&	1	&	0	&	0	&	0	&	0	\\
\text{MB7}	&	1	&	-1	&	0	&	0	&	0	&	0	&	-1	&	1	&	0	\\
\text{MB8}	&	1	&	0	&	-1	&	-1	&	0	&	1	&	0	&	0	&	0	\\
\text{MB9}	&	1	&	0	&	-1	&	0	&	0	&	0	&	-1	&	0	&	1	\\
\end{array}
\end{equation}
\end{linenomath*}
It is a straightforward exercise to confirm that we can transition between the two solutions in (\ref{eq:twosol}) by adding or subtracting moves from (\ref{eq:MBcont}) one-at-a-time without encountering a negative count. More importantly, the moves in (\ref{eq:MBcont}) can be used to connect any two solutions in the same fiber, no matter what value of $\bm y$ is observed.

\mrsadd{There is often a need to analytically find a Markov basis for a given problem.  Even though tools like {\tt 4ti2} are freely available, computation of Markov bases remains challenging.  As we discuss later, for many of the capture-recapture examples we have explored, {\tt 4ti2} can fail to compute Markov bases for studies with a moderate to large number of sampling occasions.  As we know of no simple test to confirm whether a specified set of moves $\mathcal B$ is a Markov basis, we often need to rely on theoretically derived Markov bases to confirm that our MCMC algorithms are valid.  In the following section we find such a theoretical result for a class of capture-recapture models including \Mta.}

\section{Model M$_{\mathrm{t}\alpha}$ and Simple Corruptions}
\label{sect:mta}

Here, we examine model \Mta, the specific model of misidentification considered by \Link. We fit this model into a larger class of models in which any identification error results in what we refer to as a simple corruption.  We then show that for any model in this class, we can construct a lattice basis that is guaranteed to connect every element of the fiber, irrespective of $\bm y$, i.e. it is also a Markov basis.  

Model \Mta builds on the standard closed population model with time-dependent capture probabilities, model \Mt of \citet{Otis1978}, by allowing for individuals to be misidentified when captured. The model assumes that all errors are unique meaning that an individual cannot be identified as another individual and the same error cannot occur multiple times. The result is that an error on the $j^{th}$ capture occasion leads to a ghost observed history containing a single observation on the $j^{th}$ occasion.

For this model, the vector of \mrsadd{summary} statistics, $\bm y$, contains the counts of the $2^K-1$ observable capture histories. The vector of latent variables contains the counts of the possible true histories constructed from the events:
\begin{itemize}
\item 0 -- the individual was not captured,
\item 1 -- the individual was captured and correctly identified,
\item 2 -- the individual was captured and incorrectly identified.
\end{itemize}
For example, for a study with $K=5$ capture occasions the true history $01221$ would generate three observed histories: $01001$, $00100$, and $00010$. Including the null history $0\ldots0$, the vector of true counts has length $3^K$. The configuration matrix, $\bm A$, has dimension $(2^K-1)\times3^{K}$ and $A_{ij}=1$ if the $j^{th}$ true history generates the $i^{th}$ observed history and is equal to zero otherwise. For example, the column corresponding to the history $01221$ would contain three non-zero entries in the rows associated with the observable histories $01001$, $00100$, and $00010$.  \mrsadd{A description of the model} along with the vectors $\bm x$ and $\bm y$ and matrix $\bm A$ for $K=2$ are given in the supplementary materials, with more details in \Link.

A feature of \Mta is that whenever an error in identification occurs, it involves
only one \mrsadd{individual} and results in one or more observed histories.
We define such an error as a simple corruption.
For example, the errors in true history $01221$ above affect no other true history and lead to three observed histories.
Another example of simple corruptions are the errors that occur when multiple marks cannot be matched, as described in \citet{Bonner2013a} and \cite{McClintock2013}. Suppose that a study uses photographs to identify individuals and that photographs taken from the left or right side cannot be matched without further information. In this case, any individual that is photographed from both the left and right sides on different occasions will contribute two histories to the observed data set. Using the events $L$ and $R$ to denote photographs from the left and right, the true history $0LRRL$ would generate observed histories $0L00L$ and $00RR0$. In this case, each true history will contribute one or two histories to the observed data set.

For a model that contains only simple corruptions, we have the following theorem:

\begin{mythm}
Suppose that: (i) $\bm A$ contains only the values 0 and 1 
 and (ii) the columns of $\bm A$ contain all of the columns of the identity matrix.
Then there exists a lattice basis $\mathcal L=\{\bm v_1,\ldots,\bm v_m\}$, which is also a Markov basis.
\end{mythm}
\mrsadd{The first condition (values of 0 and 1) occurs under the assumption of simple corruption,} while the second condition (columns of the identity matrix) \mrsadd{occurs when} every observable history is also a true history in which there is no misidentification.  
Provided these assumptions hold, 
then we can use the algorithm in Figure \ref{fig:alg} with a suitable lattice basis $\mathcal L$ and connect the fiber.  The proof of this theorem is provided in the supplementary materials, 
along with a description of how to \mrsadd{construct} the lattice (Markov) basis $\mathcal L$.  

\mrsadd{The conditions of Theorem 1 are satisfied for model \Mta, so that} for $K=2$ we obtain the Markov basis in (\ref{eq:MBmta})
\begin{linenomath*}
\begin{equation}\label{eq:MBmta}
\arraycolsep=0.7em
\def\arraystretch{0.7}
\begin{array}{r|rrrrrrrrr}
& x_{00} & x_{01} & x_{02} & x_{10} & x_{11} & x_{12} & x_{20} & x_{21} & x_{22}\\\hline
\text{MB1}	&	1	&	0	&	0	&	0	&	0	&	0	&	0	&	0	&	0	\\
\text{MB2}	&	0	&	-1	&	1	&	0	&	0	&	0	&	0	&	0	&	0	\\
\text{MB3}	&	0	&	-1	&	0	&	-1	&	0	&	1	&	0	&	0	&	0	\\
\text{MB4}	&	0	&	0	&	0	&	-1	&	0	&	0	&	1	&	0	&	0	\\
\text{MB5}	&	0	&	-1	&	0	&	-1	&	0	&	0	&	0	&	1	&	0	\\
\text{MB6}	&	0	&	-1	&	0	&	-1	&	0	&	0	&	0	&	0	&	1	\\
\end{array}
\end{equation}
\end{linenomath*}
The basis in (\ref{eq:MBmta}) is identical to that presented by \Link for model \Mta when $K=2$.  

\mrsadd{
The approach of \Link to finding a basis involves 
choosing pivotal (or constraining) variables when solving the set of equations $\bm A \bm x = \bm 0$ (a full description is available either in \Link, pg 180--181, or in the supplementary materials).
%
\Link chose specific pivotal variables ($x_{01}$, $x_{10}$ and $x_{11}$) when finding the basis for model \Mta when $K=2$.  However, it was implied that this choice was arbitrary and no guidance was given as to how to select pivotal variables when $K > 2$.  It turns out that changing the pivotal variables can lead to different sets of basis vectors which may not be Markov bases.  We show in the supplementary materials that for $K=2$ and a different set of pivotal variables, $x_{22}$, $x_{20}$ and $x_{11}$, the resulting basis differs from that in (\ref{eq:MBmta}).  We also show that when the conditions of Theorem 1 are satisfied, there is a specific choice of pivotal variables guaranteed to return the Markov basis $\mathcal L$.  In particular, if we order $\bm x$ as in \Link for model \Mta and take the variable corresponding to the leading non-zero entry in each row of $\bm A$ as pivotal (as was done by \Link for $K=2$), the basis found will be the Markov basis $\mathcal L$.
}

%

\mrsadd{Theorem 1 ensures} that there is at least one lattice basis which is also a Markov basis \mrsadd{for model \Mta.  However,} it does not imply that every lattice basis is a Markov basis.  For model \Mta and $K=2$ another lattice basis (found by hand) is given in (\ref{eq:LBmta})
\begin{linenomath*}
\begin{equation}\label{eq:LBmta}
\arraycolsep=0.7em
\def\arraystretch{0.7}
\begin{array}{r|rrrrrrrrr}
& x_{00} & x_{01} & x_{02} & x_{10} & x_{11} & x_{12} & x_{20} & x_{21} & x_{22}\\\hline
\text{LB1}	&	1	&	0	&	0	&	0	&	0	&	0	&	0	&	0	&	0	\\
\text{LB2}	&	0	&	-1	&	1	&	0	&	0	&	0	&	0	&	1	&	-1	\\
\text{LB3}	&	0	&	0	&	1	&	0	&	0	&	0	&	1	&	0	&	-1	\\
\text{LB4}	&	0	&	0	&	0	&	1	&	0	&	0	&	-1	&	0	&	0	\\
\text{LB5}	&	0	&	0	&	-1	&	0	&	0	&	1	&	-1	&	1	&	-1	\\
\text{LB6}	&	0	&	0	&	0	&	0	&	0	&	0	&	0	&	1	&	-1	\\
\end{array}
\end{equation}
\end{linenomath*}
Suppose the observed data are $\bm y = (363,22,174)$ (as in \Link), then two elements in the fiber are $\bm x_1 = (0,363,0,22,174,0,0,0,0)^{\prime}$ and $\bm x_{2} = (0,361,2,22,174,0,0,0,0)^{\prime}$.  We are unable to move between these two using LB1 -- LB6 in (\ref{eq:LBmta}) as moves in the algorithm in Figure \ref{fig:alg}.  In particular, if we start at (the observed history) $\bm x_{1}$ the moves LB2, LB3, LB5 and LB6 will lead to automatic rejections because they will always propose a negative value.  This means that $\bm x_1$ and $\bm x_2$ are not connected and thus the fiber is not connected.

\begin{figure}[!htbp]
\centering
\includegraphics[width=0.7\textwidth]{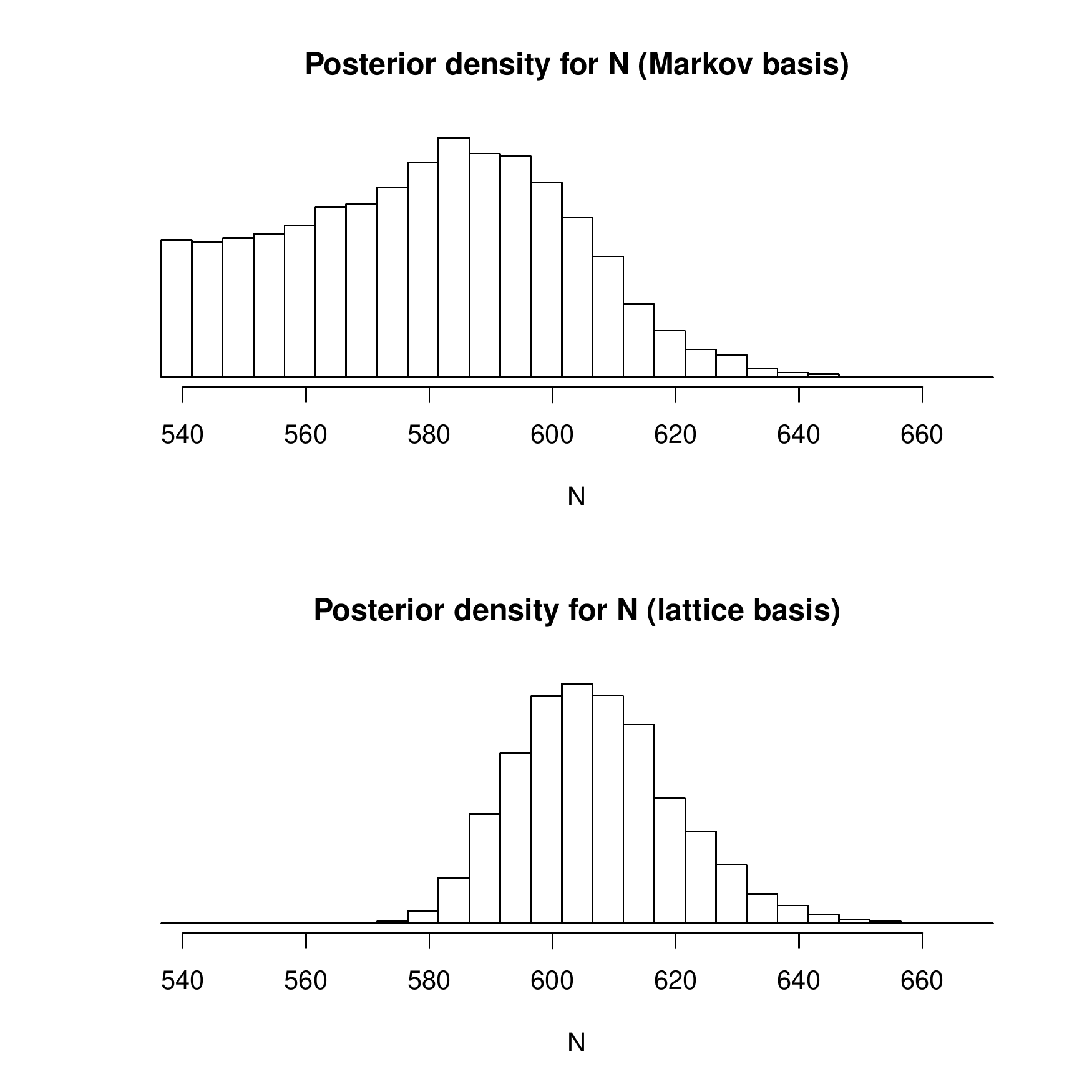}
\caption{Histograms of the estimated posterior density of $N|y$ when using the Markov basis from (\ref{eq:MBmta}) (top) and the lattice basis from (\ref{eq:LBmta}) (bottom) when starting from $\bm x_1$.}
\label{fig:postNmta}
\end{figure}

We \mrsadd{repeated} the analysis of \Link using both the Markov basis in (\ref{eq:MBmta}) and the lattice basis in (\ref{eq:LBmta}) \mrsadd{using the same prior distributions as in \Link (we used only one of the priors \Link considered for $\alpha$; a beta distribution with parameters $19$ and $1$)}.  In both cases we \mrsadd{implemented} the algorithm in Figure \ref{fig:alg} using $\bm x_1$ as the starting value \mrsadd{with interest in the abundance $N$}.  We checked convergence via trace plots and plotted the resulting distribution for $N|y$ in both cases (Figure \ref{fig:postNmta}).  The lattice basis in (\ref{eq:LBmta}) leads to a distribution for $N$ that is substantially different from the true posterior distribution and could lead to incorrect decision making.

\mrsadd{We note that efficiency gains can be made if there are observable histories with zero count.  In particular, we can delete the entries in $\bm y$ and the rows of $\bm A$ corresponding to the zero counts before deleting any columns of $\bm A$ and corresponding entries of $\bm x$ that are known to have zero count.  Provided the assumptions of Theorem 1 are still satisfied by the resulting configuration matrix then we can still find a set of moves guaranteed to connect all elements in the fiber.  The resulting set of moves is no longer a Markov basis but a Markov subbasis \cite[]{Chen2006} as it is only valid for the observed $\bm y$.  This corresponds to the approach taken by both \cite{Bonner2013a} and \cite{McClintock2013} for data with multiple marks that could not be matched.} 

This \mrsadd{section} shows that we must take care even with simple corruptions to ensure that the lattice basis we are using is also a Markov basis. 
The following two sections give examples where we do not have simple corruptions (in one of these it does not even make sense to think of corruptions in the sense of model \Mta) and a Markov basis has greater cardinality than a lattice basis.

\section{Example: Sufficient Statistics}\label{sect:summary}
Next we consider the problem of modeling data from a closed population when sufficient statistics from one or more models are provided in place of the raw data.  The raw data may not be available for a variety of reasons, e.g. privacy concerns.  
Here we assume that the population is closed and that we have the sufficient statistics associated with three commonly used models \Mt, \Mb and \Mh \cite[]{Otis1978}.
From model \Mh we have the statistics $f_{1},\ldots,f_{K}$, where $f_j$ is the number of individuals who were caught $j$ times from a total of $K$ sampling occasions; from model \Mt we have the statistics $n_{1},\ldots,n_{K}$, where $n_{j}$ is the number of individuals captured in the $j$th sample; and from model \Mb we have the statistic $M_{\cdot} = \sum_{j=1}^{t}M_{j}$, with $M_{j}$ the number of marked individuals in the population in sample $j$.  Note that we do not include the other sufficient statistics for model \Mt and \Mb noted by \cite{Otis1978} as they are deterministic functions of $f_{1},\ldots,f_{K}$.

All of these statistics are linear functions of the data which means that this problem can be expressed using the linear constraint in (\ref{eq:lincon}). In this example, $\bm x$ represents the vector of counts for the $2^K-1$ true histories; $\bm y$ represents the vector of counts for the $2K+1$ sufficient statistics; and the configuration matrix, $\bm A$, is a $(2K+1) \times (2^K-1)$ matrix. Details of how to find $\bm A$ along with an example for a study with $K=4$ occasions are provided in the supplementary materials.

Here we explore this scenario using multi-list data from a South Auckland, New Zealand, diabetes study from the Ph.D. research of \cite{Huakau2001} and included in the Ph.D. research of \cite{Sutherland2003}.  We ignore the potential errors in matching individuals between lists and assume that each individual is correctly matched (\mrsadd{see \cite{Lee2002} for how} such errors could also be \mrsadd{accounted for using the} linear constraint (\ref{eq:lincon})).  There are $K=4$ lists: general practitioners records (G), pharmacy records (P), outpatient records (O) and inpatient discharge records (D) that we assume are ordered as written.  We use the data for males and reduce the full data (which is available in \citealt{Sutherland2003}) to the statistics: $\bm n = (n_{G}, n_{P}, n_{O},n_{D})^{\prime} = (629,622,6279,1623)^{\prime}$, $\bm f = (f_{1}, f_{2}, f_{3}, f_{4})^{\prime} = (6030,1312,161,4)^{\prime}$ and $M_{\cdot} = 8680$ to give
\begin{linenomath*}
\[
\bm y = (6030,1312,161,4,629,622,6279,1623,8680)^{\prime}.
\]
\end{linenomath*}
As well as $\bm y$ being sufficient for models \Mt, \Mh and \Mb, it is also sufficient for the two-factor quasi-symmetric version of model \Mth that is induced by a Rasch model \cite[see][for details of this model]{Agresti1994}.

\mrsadd{The vector $\bm x$ is indexed by $\bm \omega = (\omega_{G},\omega_{P},\omega_{O},\omega_{D})$, where $\omega_{j}=1$ denotes inclusion on list $j$ with $\omega_{j}=0$ otherwise, so that $x_{1101}$ is the number of individuals  on lists G, P and D and not on list O. } 
Our focus here is to attempt to make inference about $x_{1000}$, 
the number of individuals who appear only in list $G$.   We may also wish to fit a model to $\bm x$ for which $\bm y$ are not sufficient statistics.  By definition, the resulting model would be nonidentifiable, but this does not necessarily mean that there is no information about parameters of this model, including the abundance $N$.  The latent multinomial model can be used in either of these situations.


A lattice basis found using the Hermite normal form is
\begin{linenomath*}
\[
\arraycolsep=0.25em
\def\arraystretch{0.7}
\begin{array}{r|rrrrrrrrrrrrrrr}
 & x_{0001} & x_{0010} & x_{0011} & x_{0100} & x_{0101} & x_{0110} & x_{0111} & x_{1000} & x_{1001} & x_{1010} & x_{1011} & x_{1100} & x_{1101} & x_{1110} & x_{1111} \\\hline
\text{LB1}	&	0	&	0	&	0	&	0	&	0	&	0	&	0	&	0	&	-1	&	0	&	1	&	1	&	0	&	-1	&	0	\\
\text{LB2}	&	0	&	0	&	0	&	0	&	0	&	0	&	0	&	0	&	-1	&	1	&	0	&	0	&	1	&	-1	&	0	\\
\text{LB3}	&	0	&	-1	&	1	&	0	&	0	&	0	&	0	&	1	&	-1	&	0	&	0	&	0	&	0	&	0	&	0	\\
\text{LB4}	&	1	&	-2	&	0	&	1	&	0	&	0	&	0	&	0	&	0	&	0	&	1	&	0	&	-2	&	1	&	0	\\
\text{LB5}	&	1	&	-2	&	0	&	0	&	1	&	0	&	0	&	1	&	-1	&	0	&	1	&	0	&	-2	&	1	&	0	\\
\text{LB6}	&	1	&	-2	&	0	&	0	&	0	&	1	&	0	&	1	&	-1	&	0	&	1	&	0	&	-1	&	0	&	0	\\
\text{LB7}	&	1	&	-2	&	0	&	0	&	0	&	0	&	1	&	1	&	0	&	0	&	0	&	0	&	-2	&	1	&	0	\\
\end{array}
\]
\end{linenomath*}

Using the seven moves LB1 -- LB7 in the algorithm in Figure \ref{fig:alg} it is impossible to move between the two solutions $\bm x_1$ and $\bm x_2$
\begin{linenomath*}
\begin{align*}
\bm x_1 &= \left(652, 4865, 794, 253, 18, 234, 62, 260, 26, 221, 67, 19, 0, 32, 4\right)^{\prime}\\
\bm x_{2} &=  \left(684, 4901, 694, 253, 31, 154, 161, 192, 49, 365, 0, 19, 0, 0, 4\right)^{\prime}.
\end{align*}
\end{linenomath*}
If we are currently at $\bm x_2$, it is clear that all moves (except LB3) will lead to at least one negative cell count and will be automatically rejected.  The vector LB3 can be used to update $\bm x_2$, but we are unable to get to $\bm x_1$ using LB3 alone.
Again, we have at least two sets of elements in the fiber that we can move within, but are unable to move between.

A Markov basis for this problem can be constructed in {\tt 4ti2} and is made up of the 16 elements given in the supplementary materials.
%
%
\mrsadd{Since (i) {\tt 4ti2} finds a minimal Markov basis, and (ii) the cardinality of the Markov basis is larger than that of a lattice basis, we can be certain that} a lattice basis can never be a Markov basis for this problem.  Even though it is likely possible to construct another lattice basis that can move between $\bm x_1$ and $\bm x_2$ there will be either (i) another two elements in the fiber that are not connected, or (ii) another two elements in the fiber for a different $\bm y$ that we cannot move between with such a lattice basis.

Here we fit model \Mt and
run the algorithm in Figure \ref{fig:alg} with both the Markov basis given in the supplementary materials and the lattice basis specified above \mrsadd{(details of the model are given in the supplementary materials)}.
We make use of the factorization theorem \citep[e.g, see][pg. 276]{Casella2002} that states that a model $f(\bm x|\bm \theta)$ with sufficient statistics $\bm y$ can be expressed as
\begin{linenomath*}
\[
f(\bm x|\bm \theta) = g(\bm x | \bm y)h(\bm y|\bm \theta).
\]
\end{linenomath*}
A practical implication is that only $g(\bm x|\bm y)$ is required if interest is in a function of $\bm x$ such as $x_{1000}$, and the parameters $\bm \theta = (N,p_1,\ldots,p_K)$ need not be specified.   A related implication is that if we do choose to update $\bm \theta$ the resulting chains will converge to the correct posterior  $[\bm \theta|\bm y]$ even if \mrsadd{we} (i) do not  update $\bm x$, or (ii) update $\bm x$ using a set of moves that is unable to connect the fiber, such as the lattice basis above; provided we specify an appropriate MCMC sampler for $\bm \theta$.

\begin{figure}[!htbp]
\centering
\includegraphics[width=0.7\textwidth]{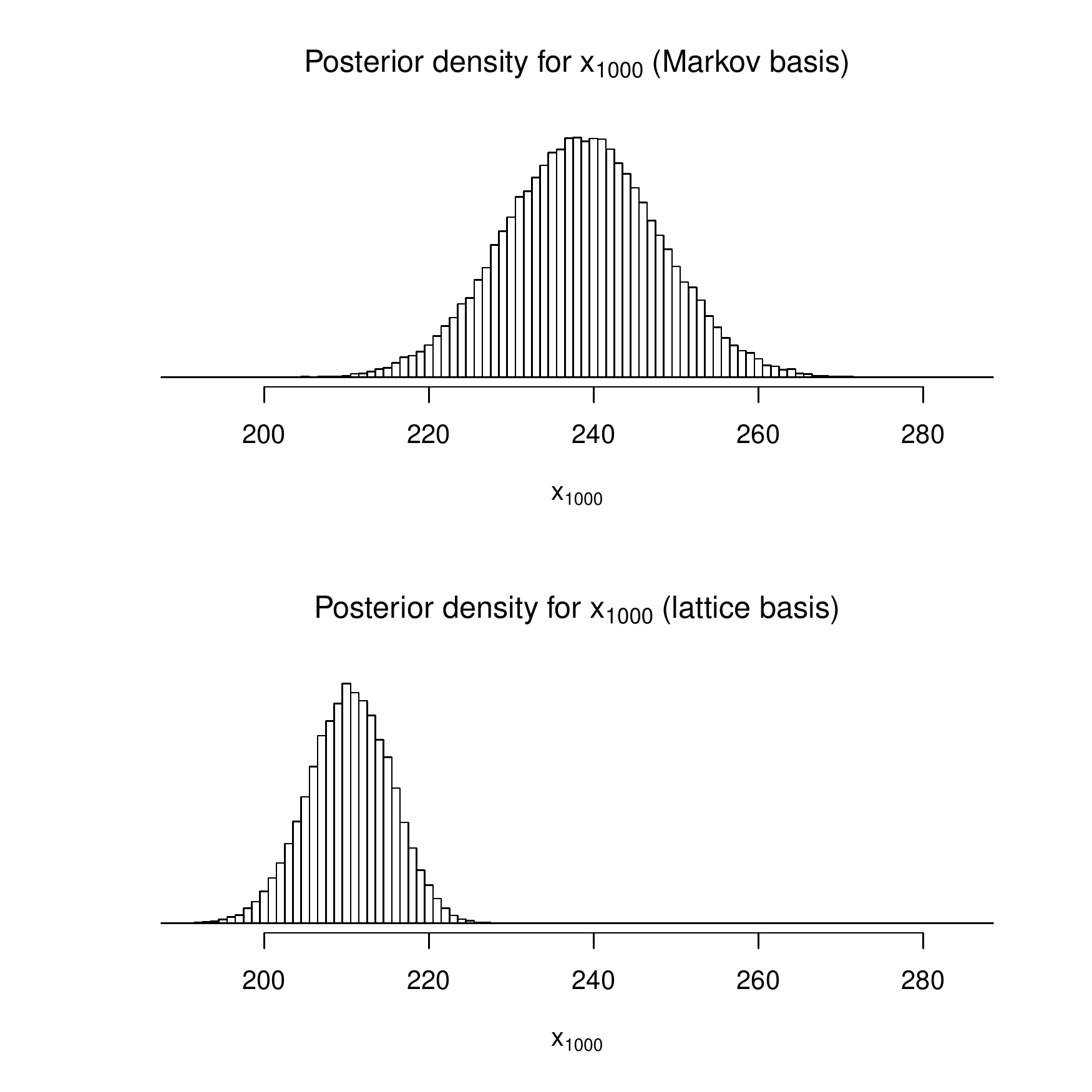}
\caption{Posterior densities of $x_{1000}$ when using the Markov basis from the supplementary materials (top) and the lattice basis specified in section 4 (bottom) when starting at $\bm x_2$.}
\label{fig:suffcdf}
\end{figure}

Using the lattice basis and starting at $\bm x_{2}$ the resulting distributions for $x_{1000}$ are qualitatively different from the posterior distribution found using the Markov basis even though the individual chains appear to have converged to the stationary distribution  (Figure \ref{fig:suffcdf}).  The true value of $x_{1000} = 260$ has some posterior mass when using a Markov basis (despite being in the tail).  If we were to believe the results when using the lattice basis $x_{1000} = 260$ is so far in the tail, we would conclude it has negligible posterior mass.

\section{Example: Band Misreading in Mark-Resight}
\label{sect:resight}

As a final example we consider a mark-resight model which allows for the possibility that individuals are \mrsadd{misidentified} when resighted in the field. Imagine that \mrsadd{there are $K_1$ distinct occasions, on which researchers capture a number of} unmarked individuals, mark them, and release them back into the population. \mrsadd{Along with that are a series of $K_2$ resighting occasions, on which} the researchers conduct visual surveys to identify previously marked individuals. Data from the experiment consist of the observed \mrsadd{resighting} histories for each individual. If there were no errors then standard mark-resight models could be used to estimate survival or movement rates \cite[e.g.][]{Hestbeck1991}; or abundance \cite[e.g.][]{McClintock2006}.

Suppose now that individuals may be misidentified when they are resighted. In direct contrast to model \Mta, which assumes that errors are unique and never match other individuals, we assume that errors may be repeated and always match the identity of previously marked individuals. The justification for this assumption is that the available set of marks is known on each occasion when individuals are identified by man-made marks instead of natural markers (e.g., genotypes or photo-id). Erroneous sightings of marks which have not been released can then be identified and removed from the data prior to the analysis. The only time an error cannot be detected and discarded is when one previously marked individual is misidentified as another previously marked individual.  \mrsadd{We note that removal of erroneous sightings is only justified when estimating survival.  Removing erronous sightings when including unmarked individuals would lead to biased estimators of abundance \cite[]{McClintock2013a}.}

\mrsadd{For the remainder of the section, we assume that the capture and resighting occasions occur simultaneously so that \(K=K_1=K_2\).} The true capture histories for each individual can now be constructed in terms of four possible events. On each occasion, individual $i$ may be:
\begin{itemize}
\item not captured or resighted (event 0),
\item captured or resighted and correctly identified (event 1), or
\item resighted and incorrectly identified (event 2).
\end{itemize}
Further to this, another individual may be resighted and incorrectly identified as individual $i$ (event 3). Events 2 and 3 represent false negative and false positive \mrsadd{resightings}. For example, the history 123 for individual $i$ would indicate that $i$ was captured and marked on the first occasion, was resighted and misidentified on the second occasion, and that another individual was resighted and identified as $i$ on the third occasion of a study with $K=3$ occasions. To simplify the example, we assume that individuals cannot be misidentified when they are first captured and that multiple events involving the same individual cannot occur on a single occasion (e.g., it is not possible to resight $i$ and incorrectly identify another individual as $i$ on the same occasion).  \mrsadd{This assumption may be unrealistic in some situations and was made to make the approach tractable.  Developing methodology to relax this assumption is ongoing research.}

For an experiment with $K$ occasions, the model has $(4^K-1)/3$ possible true histories and the usual $2^K-1$ observable histories. Further to this, there are $K-1$ extra constraints that equate the number of false negatives and false positives (2s and 3s) on occasions 2 through $K$. As a result, $\bm A$ has dimension $(2^K + K-2)\times (4^K-1)/3$ and a basis for $\ker_{\mathbb Z}(\bm A)$ has $(4^K-1)/3 - (2^K + K-2)$ elements.

To make this more concrete, we consider the specific case of an experiment comprising $K=3$ occasions. In this case, there are $(4^3-1)/3=21$ possible true histories, $2^3-1=7$ observable histories, and $3-1=2$ extra constraints on the number of false \mrsadd{positive} and negative \mrsadd{resightings} (2s and 3s) on occasions 2 and 3. Details of how to construct $\bm A$ along with $\bm x$ and $\bm y$ for a study with $K=3$ capture occasions are provided in the supplementary materials. In this case, a basis for $\ker_{\mathbb Z}(\bm A)$ has $12$ elements and the specific lattice basis obtained using the Hermite normal form is provided in the supplementary materials, along with the Markov basis, computed using \verb?4ti2?, that has $63$ elements.

To illustrate the problems that can occur with \mrsadd{t}his model we first consider the analysis of a single (fake) data set. Suppose  that each observable history is recorded one time so that
\[
\bm y = (1,1,1,1,1,1,1).
\]
An exhaustive search confirms that the fiber defined by $\bm y$ contains exactly $120$ unique elements. However, the lattice basis given in the supplementary materials does not connect all of the elements in the fiber. Instead, the lattice basis divides the fiber into two distinct pieces including a large set of 87 connected elements; and a further set of 33 isolated elements which connect to nothing else. As a result, the distribution of the sample generated by the algorithm in Figure \ref{fig:alg} using the elements of the lattice basis in the supplementary materials as moves will depend on the starting point.

\begin{figure}
  \centering
  \begin{tabular}{cc}
  \includegraphics[width=.5\textwidth]{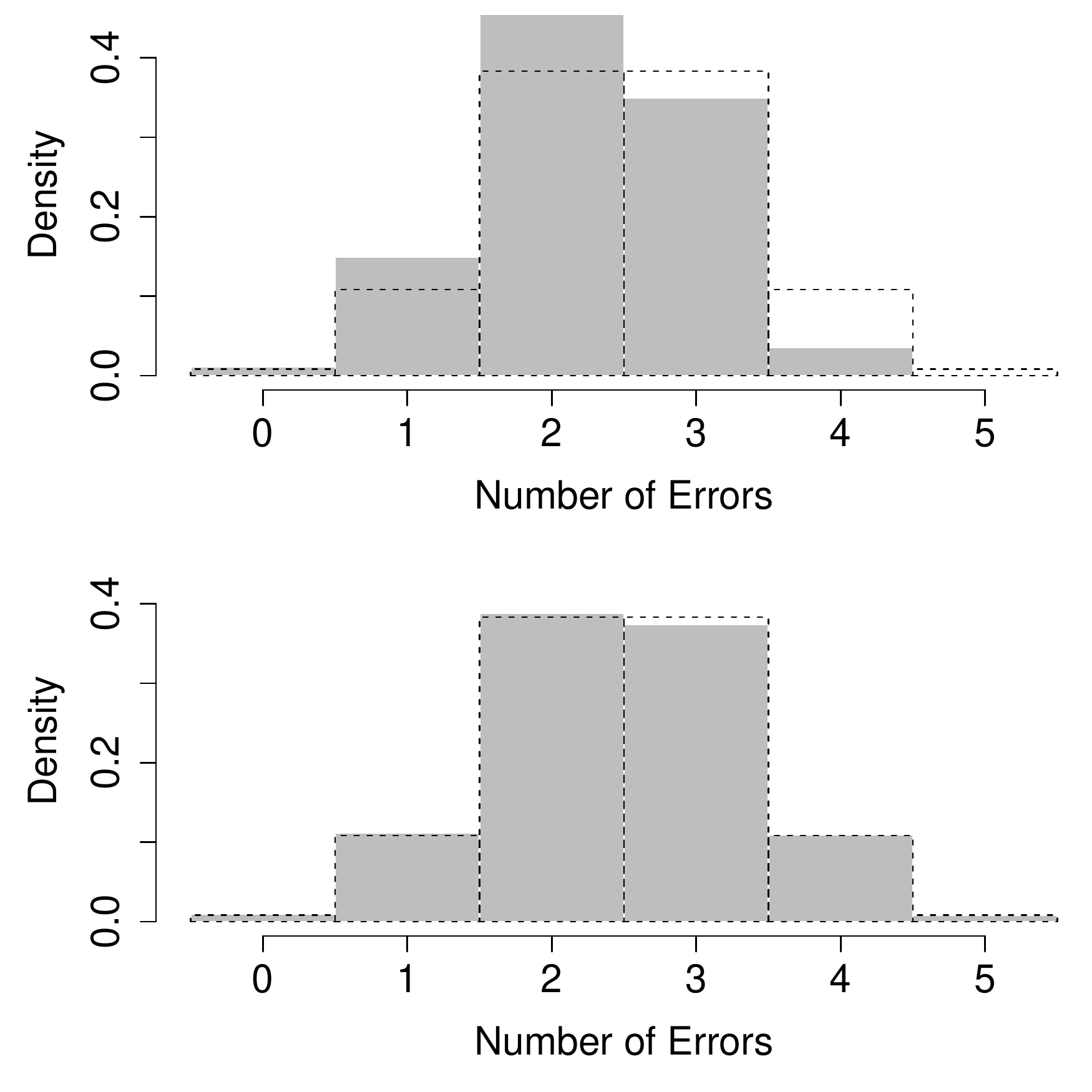}
  \end{tabular}
  \caption{Distributions of the number of errors in the solutions sampled given the data $\bm y$. The top histogram illustrates the distribution generated using the lattice basis with the starting value $\bm x_1$. The bottom plot illustrates the distribution obtained using the full Markov basis with the same starting value. In each plot, the gray bars represent the distribution of the number of errors while the dashed bars represent the true distribution over all 120 unique solutions.}
  \label{fig:bre-results-1}
\end{figure}

To show this, we have investigated the output from the algorithm in Figure \ref{fig:alg} when using a lattice basis as our set of moves.  We have chosen a starting point that lies in the largest part of the fiber and connects with 86 other elements:
\begin{linenomath*}
\[
  \bm x_1 = (1,0,1,1,0,1,1,0,0,1,0,0,1,0,0,0,0,0,0,0,0)^{\prime}. 
\]
\end{linenomath*}
\mrsadd{Assuming a multinomial distribution for $[\bm x|\bm \theta]$ is not appropriate to account for the band misreading process and specification of a more complex $[\bm x|\bm \theta]$ is ongoing research.  As our goal is to show that a lattice basis is unable to connect the fiber, we simplify the model by setting $[\bm x|\bm \theta] \propto 1$}. A valid sampler should then sample uniformly from the 120 elements in the fiber. For comparison, we have also run a chain using the full Markov basis starting at $\bm x_1$.  
As expected, the first chain visits 87 unique solutions and the second visits all 120. To visualize the impact this can have on inference, Figure \ref{fig:bre-results-1} compares the distributions of the number of errors in the solutions identified by each chain. Using the lattice basis, the first chain oversamples the solutions with too few errors, placing too much mass on solutions with one or two errors and not enough on solutions with three, four, or five errors. 
In comparison, the distribution generated using the full Markov basis matches the true distribution of the number of errors in the 120 elements almost exactly.

\section{Discussion}\label{sect:discuss}
Here we have presented examples of capture-recapture models that show the importance of using a Markov basis when sampling from a linearly constrained vector of counts.  In particular, we have demonstrated the danger of using elements of a lattice basis as one-at-a-time moves in an algorithm as in Figure \ref{fig:alg}.  In many situations a  set referred to as a Markov basis is needed to ensure we can move between various elements of the fiber without passing through invalid (negative) counts.  Even when a Markov basis is a lattice basis, we must take care because not every lattice basis is a Markov basis.

For a given matrix $\bm A$ the need for a Markov basis over a simpler lattice basis depends on the lattice basis chosen, as well as the data observed.  If we consider the lattice basis for the $3 \times 3$ contingency table in section \ref{sect:basis}, difficulties arose because our data had a row sum of 0.  
A related issue is that even when a lattice basis is unable to connect the fiber, it may still be able to connect nearly all elements in the fiber.  In such a case, using a lattice basis may lead to a distribution that is an acceptable approximation of the true posterior distribution.  This is especially the case if the elements of the fiber that are not connected to the initial value are in areas of low probability in the model $[\bm x|\bm \theta]$.  This can be seen in the example from Section \ref{sect:summary}: using the lattice basis and starting at the second starting value (Figure \ref{fig:twostart}; right panel) results in an estimated posterior density that is practically indistinguishable from the true posterior distribution (Figure \ref{fig:suffcdf}).  However, there is no guarantee that any given lattice basis will provide a good approximation to the fiber.  It is possible that even with multiple starting values we may choose values that only connect a small proportion of the fiber.  

One important aspect that we have only briefly mentioned is the difficulty in constructing Markov bases.
For the purposes of this manuscript we have overcome this difficulty through (i) analytical results, or (ii) the use of the software package {\tt 4ti2} \cite[]{Hemmecke2013}.  While the latter is possible for the examples we explored, it is unable to evaluate a Markov basis for some capture-recapture examples with a moderate to large number of sampling occasions.  For example, {\tt 4ti2} was unable to compute a Markov basis (on the lead authors work machine) for the band read error model in section \ref{sect:resight} for $K > 4$.  If we were to use {\tt 4ti2} for model \Mta (ignoring the theorem presented in section \ref{sect:mta}), {\tt 4ti2} was unable to compute a Markov basis for $K > 5$.
The implication of this is that for an algorithm in the spirit of Figure \ref{fig:alg} to be implemented for problems not involving simple corruptions, methodological work is likely to be necessary to ensure a potential set of moves is a Markov basis.

Several alternative algorithms and methods have been proposed for sampling from the fiber that avoid the calculation of a full Markov basis.  We anticipate that such approaches may be useful for a range of capture-recapture examples.  These include independent sampling of elements of the fiber \cite[e.g., see][]{Chen2005}, extending the algorithm to allow limited travel through vectors $\bm x$ that contain negative values while using a set of moves that is not guaranteed to connect the fiber \cite[e.g., see][]{Bunea2000} and approaches that dynamically find a Markov basis as the algorithm runs \cite[e.g., see][]{Dobra2012}.  While promising, we expect these approaches will require adapting to the particular challenges faced in problems involving misidentification in capture-recapture data.





\bibliographystyle{plainnat}
\bibliography{ref}



\end{document}